\documentclass[aps,prl,twocolumn,showpacs,10pt,superscriptaddress,preprintnumbers,nofootinbib]{revtex4-1}
\usepackage{epsfig,amssymb,amsmath,psfrag,epstopdf,color}
\pdfoutput=1
\usepackage{graphicx}
\usepackage{subcaption}
\usepackage{ragged2e}
\DeclareCaptionJustification{justified}{\justifying}
 \usepackage[normalem]{ulem}
\usepackage{paralist}
\usepackage{hyperref}
\allowdisplaybreaks

\def\beq{\begin{equation}}
\def\eeq{\end{equation}}
\def\bsp#1\esp{\begin{split}#1\end{split}}
\newcommand{\be}{\begin{equation}}
\newcommand{\ee}{\end{equation}}
\newcommand{\bea}{\begin{eqnarray}}
\newcommand{\eea}{\end{eqnarray}}

\def\Fig#1{Fig.~{\ref{#1}}}

\newcommand{\df}{\mathrm{d}}

\newcommand{\comment}[1]{}

\newcommand{\cE}{\mathcal{E}}
\newcommand{\cQ}{\mathcal{Q}}

\begin{document}

\title{Energy Correlators Taking Charge}

\preprint{MIT-CTP-5590}

\author{Kyle Lee}
\email{kylel@mit.edu}
\affiliation{Center for Theoretical Physics, Massachusetts Institute of Technology, Cambridge, MA 02139}

\author{Ian Moult}
\email{ian.moult@yale.edu}
\affiliation{Department of Physics, Yale University, New Haven, CT 06511}


\begin{abstract}
The confining transition from asymptotically free partons to hadrons remains one of the most mysterious aspects of Quantum Chromodynamics. 
With the wealth of high quality jet substructure data we can hope to gain new experimental insights into the details of its dynamics.
Jet substructure has traditionally focused on correlations, $\langle \mathcal{E}(n_1) \mathcal{E}(n_2) \cdots \mathcal{E}(n_k) \rangle$, in the energy flux of hadrons.  
However, significantly more information about the confinement transition is encoded in how energy is correlated between hadrons with different quantum numbers, for example electric charge.
In this \emph{Letter} we develop the field theoretic formalism to compute general correlations, $\langle \mathcal{E}_{R_1}(n_1) \mathcal{E}_{R_2}(n_2) \cdots\mathcal{E}_{R_k}(n_k) \rangle$,  between the energy flux carried by hadrons with quantum numbers $R_i$, by introducing new universal non-perturbative functions, which we term joint track functions. 
Using this formalism we show that the strong interactions introduce enhanced small angle correlations between opposite-sign hadrons, relative to like-sign hadrons, identifiable as an enhanced scaling of $\langle \mathcal{E}_+(n_1) \mathcal{E}_-(n_2) \rangle$ relative to $\langle \mathcal{E}_+(n_1) \mathcal{E}_+(n_2) \rangle$.
We are also able to compute the scaling of a $C$-odd three-point function, $\langle \mathcal{E}_\mathcal{Q}(n_1) \mathcal{E}_\mathcal{Q}(n_2) \mathcal{E}_\mathcal{Q}(n_3) \rangle$.
Our results greatly extend the class of systematically computable jet substructure observables, pushing perturbation theory deeper into the parton to hadron transition, and providing new observables to understand the dynamics of confinement.
\end{abstract}

\maketitle

\emph{Introduction.}---Collider experiments provide a unique window into the dynamics of Quantum Chromodynamics (QCD). One of the primary mysteries of QCD, and gauge theories more generally, is the real time dynamics of confinement, namely of the transition from weakly interacting quarks and gluons to hadrons. Clues to the nature of this transition are imprinted in subtle patterns in energy flux at colliders, but are difficult to decode due to our lack of non-perturbative field theory techniques. A better understanding of the hadronization transition will enhance all aspects of collider physics, from precision measurements, to searches for new physics.

One of the primary recent advances in collider physics has been the development of jet substructure \cite{Larkoski:2017jix,Kogler:2018hem}, which has enabled the measurement and theoretical understanding of correlations in energy flux within individual high energy jets. Theoretically, jet substructure is studied through correlation functions \cite{Basham:1979gh,Basham:1978zq,Basham:1978bw,Basham:1977iq,Dixon:2019uzg,Chen:2020vvp}, $\langle \cE(n_1) \cE(n_2) \cdots \cE(n_k) \rangle$,  of energy flow operators, $ \cE(n)$ \cite{Sveshnikov:1995vi,Tkachov:1995kk,Korchemsky:1999kt,Bauer:2008dt,Hofman:2008ar,Belitsky:2013xxa,Belitsky:2013bja,Kravchuk:2018htv}. For applications, see e.g. \cite{Komiske:2022enw,Holguin:2022epo,Lee:2022ige,Liu:2022wop,Liu:2023aqb,Cao:2023rga,Devereaux:2023vjz,Andres:2022ovj,Andres:2023xwr,Craft:2022kdo}. Due to their infrared and collinear (IRC) safety \cite{Kinoshita:1962ur,Lee:1964is}, these observables exhibit rigorous factorization theorems \cite{Collins:1989gx,Collins:1988ig,Collins:1985ue}, providing a separation of perturbative and non-perturbative physics, and enabling systematically improvable calculations in complicated collider environments \cite{Dixon:2019uzg,Lee:2022ige}. However, despite the appealing features of IRC safe observables, by their very nature they are insensitive to many of the details of the hadronization transition which one would like to understand.

To gain further insight into the parton-to-hadron transition will require extending the class of factorizable observables through the identification of new universal non-perturbative matrix elements. These can then be combined through the renormalization group (RG), with state-of-the-art perturbative calculations, allowing modern developments in perturbative field theory to be optimally utilized to explore the confinement transition.  While this has long been appreciated, it is technically difficult to achieve. The non-perturbative matrix elements typically required in factorization theorems are \emph{functions} of longitudinal momentum fractions, which has two drawbacks: first, these functions are often poorly known leading to large uncertainties and murky conclusions, and second, these functions are convolved into perturbative calculations, making higher order perturbative calculations with modern techniques intractable.
\begin{figure}
\includegraphics[width=0.44\textwidth]{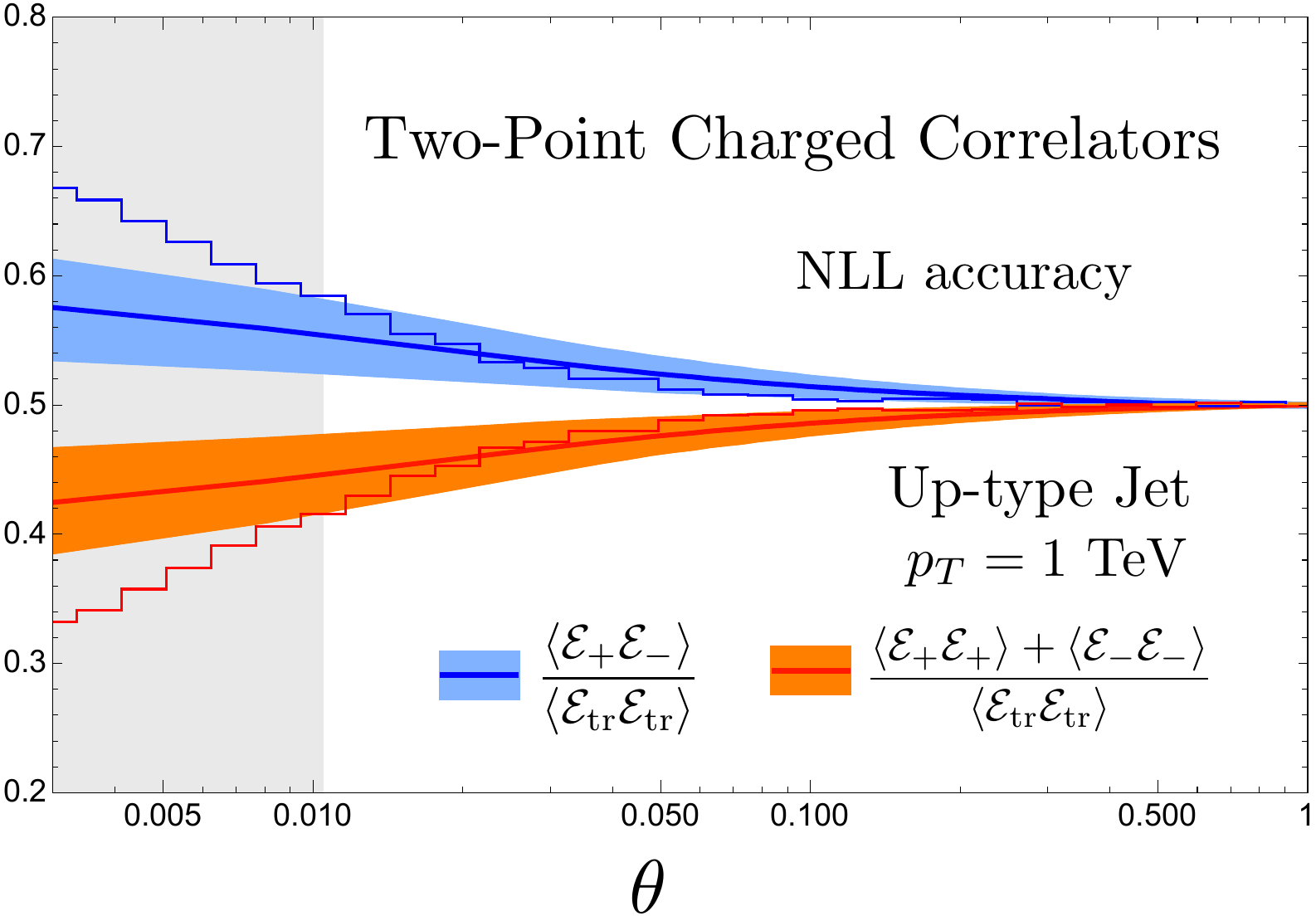}
  \caption{Scaling of two point correlators of the $\cE_+$ and $\cE_-$ detectors, showing that the strong interactions generate enhanced correlations between opposite sign hadrons as compared to like sign hadrons. Theoretical calculations at NLL accuracy are shown in solid and Pythia as a histogram. The transition to the non-perturbative regime is shown in grey.}
  \label{fig:scaling_pm}
\end{figure}

\begin{figure*}
\includegraphics[scale=0.25]{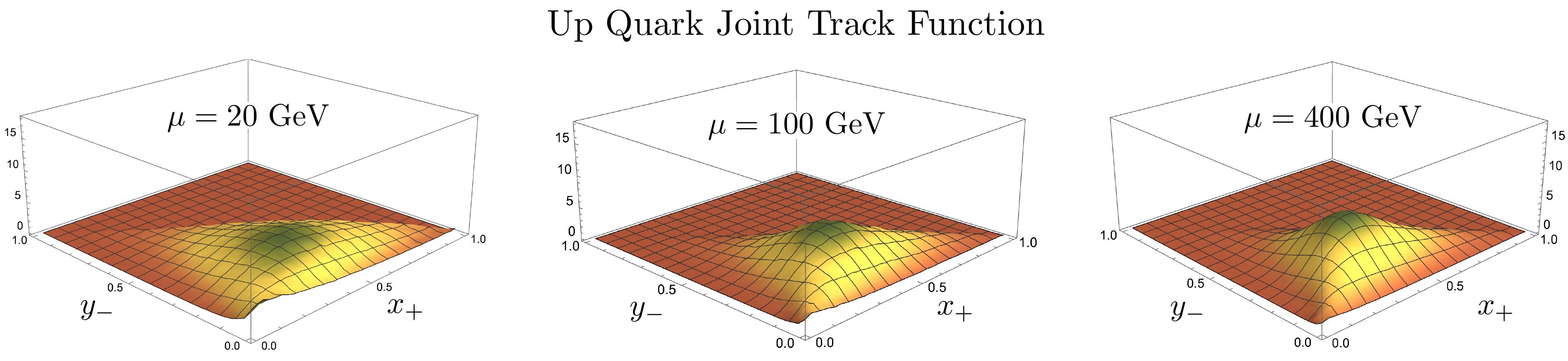} 
\caption{The up quark joint track function, $T^{+-}(x_+,y_-)$, as extracted from Pythia. The distribution shows that up type quarks deposit more of their energy into positively charged hadrons. The covariance of the distribution decreases monotonically as it is evolved to higher energies, in agreement with renormalization group evolution.}
\label{fig:joint_track}
\end{figure*} 

There has recently been progress in incorporating non-perturbative information into perturbative calculations, of the energy correlators.
Due to the correlator structure, $\langle \cE(n_1)\cdots \cE(n_k) \rangle$, of these observables, non-perturbative information enters in a fundamentally different way as compared to traditional jet shapes. 
It was shown in \cite{Chen:2020vvp} that only \emph{integer moments} of non-perturbative functions enter as matching coefficients, physically describing the relation between the perturbative and non-perturbative detectors. This has enabled progress in the theoretical description of correlations in the energy flux, $\langle \cE_{R}(n_1) \cdots \cE_{R}(n_k) \rangle$, carried by hadrons specified by a particular quantum number, $R$, \cite{Li:2021zcf,Jaarsma:2022kdd,Chen:2022pdu,Chen:2022muj} using the track function formalism \cite{Chang:2013rca,Chang:2013iba}. However, this is just the beginning of a much broader expansion of the class of calculable observables enabled by the introduction of the energy correlators.

The most general energy correlator consists of the correlation between energy flux carried by different quantum numbers, $\langle \cE_{R_1}(n_1) \cE_{R_2}(n_2) \cdots \cE_{R_k}(n_k) \rangle$. Such correlators allow one to ask many questions about  the hadronization transition that are inaccessible to standard IRC safe observables. For example, one may wonder if the strong interactions produce more small angle correlations between like-sign or same-sign hadrons, as characterized by the two-point correlator  $\langle \cE_+(n_1) \cE_-(n_2) \rangle$. Such correlations could be introduced microscopically through string fragmentation \cite{Andersson:1983ia}, for example. Observables with similar physics goals were introduced in \cite{Chien:2021yol} based on correlations of leading hadrons. However, such observables are complicated theoretically, and difficult experimentally due to possible re-orderings of leading hadrons due to decays. Motivated by the recent progress combining track functions and energy correlators, one can ask whether such quantities can be computed in a systematic fashion using universal non-perturbative inputs, and if so, what is this structure of those inputs?

In this \emph{Letter}, we show that we can systematically compute correlation functions of the form $\langle \cE_{R_1}(n_1) \cE_{R_2}(n_2) \cdots \cE_{R_k}(n_k) \rangle$, by introducing a new class of universal non-perturbative functions characterizing the fragmentation process. We show how the moments of these functions enter factorization theorems for these generalized correlators, and we apply them to show that the hadronization process of the strong interactions introduces enhanced correlations for like sign hadrons at small angles, namely that $\langle \cE_+(n_1) \cE_-(n_2) \rangle$ exhibits an enhanced scaling at small angles relative to $\langle \cE_+(n_1) \cE_+(n_2) \rangle$, as shown in \Fig{fig:scaling_pm}.

\emph{Generalized Detectors.}---Collider measurements are made at asymptotic infinity, i.e. in the deep infrared (IR) of the theory. The space of field theoretically well defined detectors has primarily been studied in the context of conformal field theories (CFTs) (see \cite{Hofman:2008ar,Belitsky:2013xxa,Belitsky:2013bja,Kravchuk:2018htv,Caron-Huot:2022eqs}), for which there is no distinction between the IR and the ultraviolet (UV). In this case, the detectors typically measure conserved charges of continuous symmetries, and can be expressed as operators composed of the UV fields. These include detectors for Lorentz charges, such as the standard energy flow operator \cite{Sveshnikov:1995vi,Tkachov:1995kk,Korchemsky:1999kt,Bauer:2008dt,Hofman:2008ar,Belitsky:2013xxa,Belitsky:2013bja,Kravchuk:2018htv}
\begin{align}
    \cE(\vec n_1) = 
    \lim_{r\rightarrow \infty} \int \mathrm{d}t \,r^2 n_1^i \,
    T_{0i}(t,r\vec{n}_1)\,,
    \label{eq:def}
\end{align}
or any internal U$(1)$ symmetry \cite{Hofman:2008ar,Belitsky:2013xxa,Belitsky:2013bja}, for example electromagnetic charge,
\begin{align}
    \cQ(\vec n_1) = 
    \lim_{r\rightarrow \infty} \int \mathrm{d}t \,r^2 n_1^i \,
    J_{i}(t,r\vec{n}_1)\,.
    \label{eq:def}
\end{align}
Correlators of $\cE$ are infrared and collinear safe, while correlators of $\cQ$ suffer from infrared divergences in gapless theories. See \cite{Chicherin:2020azt} for a calculation of $\cQ$ correlators.

However, since QCD in the IR is a gapped theory of non-interacting hadrons, correlations of any properties of the hadrons can be measured. However, generic observables are hopelessly non-perturbative, making it difficult to gain any insights into their dynamics by leveraging perturbative techniques.
In this \emph{Letter}, we will extend beyond IRC safe observables, by constructing generalized detectors for which  the non-perturbative corrections associated to the detector are a universal, i.e. process independent, property of the detector itself. From the perspective of perturbative QCD,  this implies that we want to study detectors where the associated divergences are collinear in nature (as opposed to multiplicity and charge which have \emph{soft} divergences).  

A general way of creating such detectors in a gauge theory is to ensure that they always have a positive power of energy (i.e. guaranteeing $J> 1$ \cite{Kologlu:2019mfz}). There are two ways of doing this: we can either consider products of the energy flow operator with a charge, which we denote $\cE_{\cQ}$, or we can consider the energy flux restricted to some subset of hadrons $\cE_R$.  The action of the operator $\cE_{\cQ}$ is defined by $\cE_{\cQ}(\vec n_1)|k\rangle=E_k Q_k \delta(\hat n_1-\hat k) |k \rangle$, which involves multiplication of the energy with the charge. It is a $C$-odd detector, acquiring a sign under charge conjugation.
It is perhaps more appropriate to denote it $\cE\cQ$, but we prefer the condensed notation. In the first case, we believe that the detectors will be well defined in a generic theory (e.g. also in a CFT) after renormalization \cite{Caron-Huot:2022eqs}. The second case is of interest when we have a gapped non-interacting theory in the IR, so that $\cE=\sum_i \cE_i$, with $i$ the hadron type. In this case, which is relevant for real world QCD, it is well defined to measure the energy flux carried by hadrons of a specific type (or any collection of hadrons specified by some property such as positive electromagnetic charge). We do not believe that such detectors are defined in generic gapless theories.

Motivated by these considerations, we are led to introduce two general classes of charged (energy) correlators: $\langle \cE_{R_1}(n_1) \cdots \cE_{R_k}(n_k) \rangle$, where the $R_i$ are restricted sets of hadronic states, and $\langle \cE_{\cQ_i}(n_1) \cdots \cE_{\cQ_k}(n_k) \rangle$, where $Q_i$ are a collection of U$(1)$ charges. We believe that these provide the appropriate generalization of energy correlators to the study of correlations of charge at collider experiments. We will derive factorization theorems for both these classes of correlators, and show that they depend on moments of a new universal non-perturbative input. In this \emph{Letter} we will focus on the two point functions $\langle \cE_+(n_1) \cE_-(n_2) \rangle$ and  $\langle \cE_\cQ(n_1) \cE_\cQ(n_2) \rangle$, and their applications to the confinement transition, as well as the interesting $C$-odd three-point function $\langle \cE_\cQ(n_1) \cE_\cQ(n_2) \cE_\cQ(n_3) \rangle$. We will consider the more general case of $N$-point functions in a companion paper.

\begin{figure}
\includegraphics[width=0.40\textwidth]{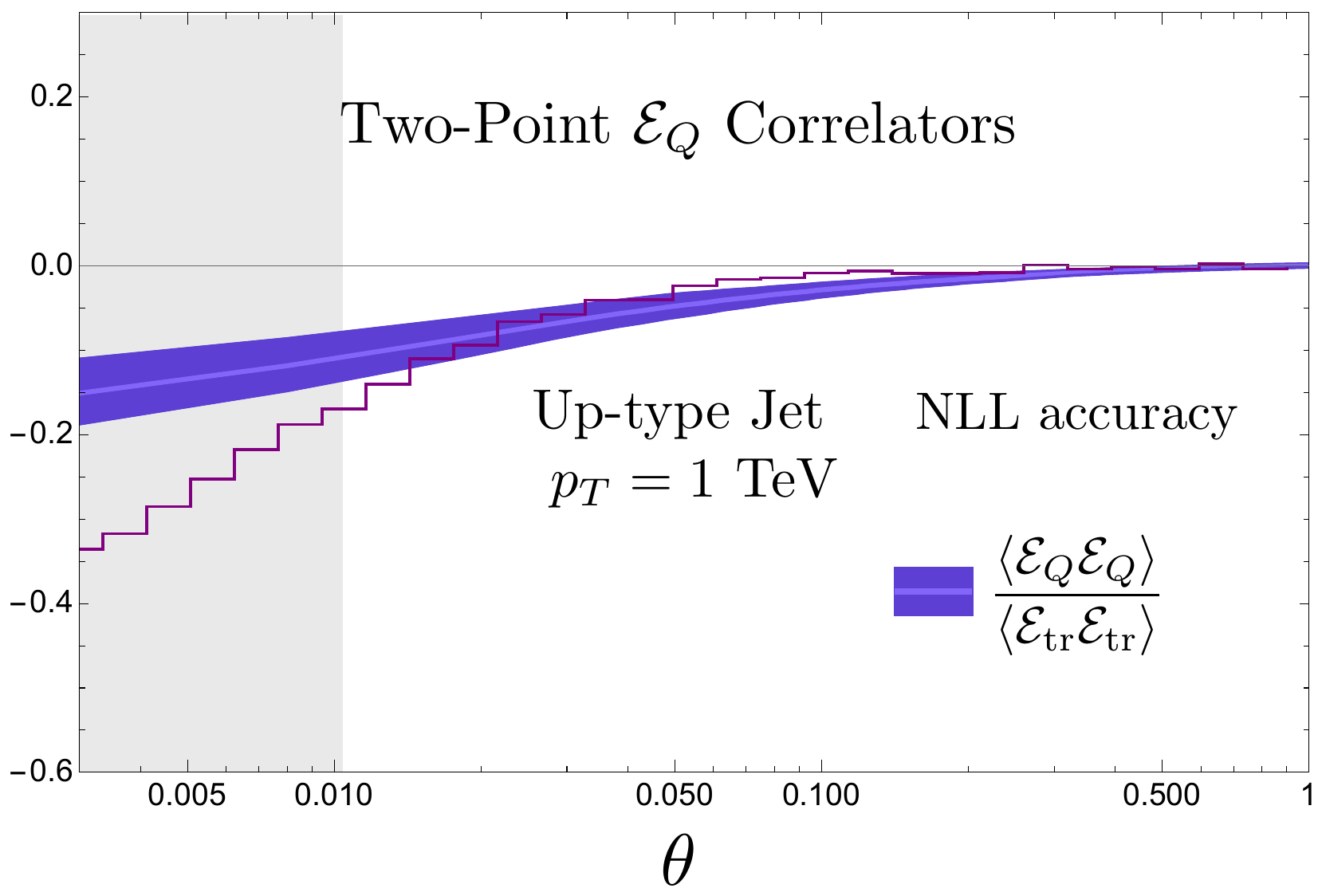}\\
\includegraphics[width=0.40\textwidth]{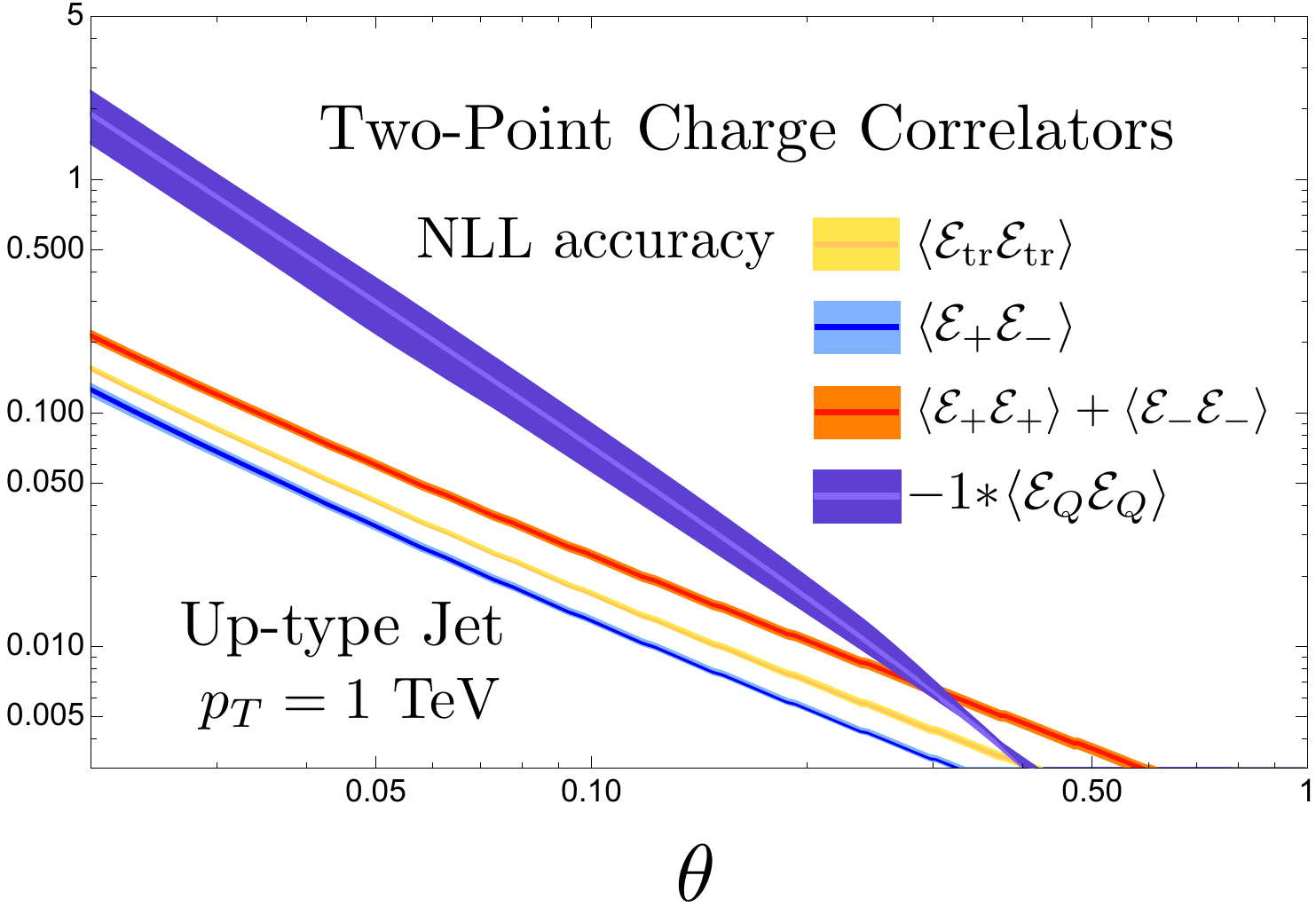}
  \caption{The small angle limit of the $\langle \cE_Q \cE_Q \rangle$ correlator for up quark jets plotted in both linear-log and log-log. Analytic calculations at NLL accuracy are shown in solid, and Pythia is shown as a histogram. The transition to the non-perturbative region is shown in grey. The inclusion of $\cQ$ qualitatively modifies the power law as compared to a standard two-point correlator of energy flow operators.}
  \label{fig:scaling_EQ}
\end{figure}

\emph{Joint Track Functions.}---To derive factorization theorems for these more general correlators requires the introduction of new universal non-perturbative functions, which we call ``joint track functions". For the specific case of two charges, which we denote $+$, $-$, the quark joint track function is defined as
\begin{align} \label{T_def}
&T_{q}^{+-}(x_+,y_-)=\!\int\! \df y^+ \df ^{d-2} y_\perp  \sum_X  \delta \biggl( x\!-\!\frac{P_{+}^-}{k^-}\biggr)  \delta \biggl( y\!-\!\frac{P_{-}^-}{k^-}\biggr)\nonumber \\
&\hspace{-0.2cm}\cdot e^{ik^- y^+/2} \frac{1}{2N_c}
\text{tr} \biggl[  \frac{\gamma^-}{2} \langle 0| \psi(y^+,0, y_\perp)|X \rangle \langle X|\bar \psi(0) | 0 \rangle \biggr]\,.
\end{align}
Here $P_{+}^-$ and  $P_{-}^-$ denote the momentum fraction carried by all positive or negative hadrons in the state $X$, respectively. We have suppressed gauge links in the definition for compactness. The gluon joint track function is defined in a similar manner, and the extension to generic charges, and their full RG structure, will be presented in a companion paper. As compared to track functions \cite{Chang:2013rca,Chang:2013iba}, the joint track function simultaneously measures the energy fractions carried by hadrons of multiple distinct quantum numbers. We restrict ourselves to positive and negative electric charges for this paper.

We have extracted the joint track function using the Pythia parton shower \cite{Sjostrand:2014zea,Sjostrand:2007gs}, which implements a string model of fragmentation \cite{Andersson:1983ia}. Plots for several different values of the RG scale $\mu$, are shown in \Fig{fig:joint_track} for up type quarks. We clearly see that up type quarks fragment more of their energy into positively charge hadrons, as compared to positively charged hadrons. As the distribution evolves towards higher energies, its covariance decreases monotonically, as will be derived from the RG.

The joint track functions are complicated multi-variable non-perturbative functions. However, correlators are only sensitive to specific \emph{moments} of the track functions, enabling predictive power.  To this end, we define the joint track moments as
\begin{align}
&T_i^{+-}(n,m;\mu) =\int_0^1\hspace{-0.2cm} \,dx\,dy\, x^n\,y^m\, T_i^{+-}(x,y;\mu)\,, \nonumber
\end{align} and the asymmetric moments as
\begin{align}
T_i^{\mathrm{asy}}(N;\mu)=&\int_0^1 dx\, dy\, (x-y)^N\, T_i^{+-}(x,y;\mu)\,.
\end{align}
We will see that the two-point correlator $\langle \cE_+(n_1) \cE_-(n_2) \rangle$ is sensitive to $T^{+-}(1,1;\mu)$, while  $\langle \cE_Q(n_1) \cE_Q(n_2) \rangle$ is sensitive to $T^{\mathrm{asy}}(2;\mu)$.

Much like for fragmentation functions, the RG evolution of joint track functions can be computed in perturbation theory. A key advance in jet substructure has been the ability to compute RG equations that go beyond the traditional \emph{linear} DGLAP \cite{Dokshitzer:1977sg,Gribov:1972ri,Altarelli:1977zs} paradigm.
The RG evolution of the joint track functions can be derived from that for the standard track functions \cite{Li:2021zcf,Jaarsma:2022kdd,Chen:2022pdu,Chen:2022muj}. For the particular case of the second moments of joint track functions of two charges, we have
\begin{align}
\frac{\mathrm{d}}{\mathrm{d} \ln \mu^2} \vec{\sigma}^{+-}(1,1;\mu)  &=-\hat{\gamma}(3) \vec{\sigma}^{+-}(1,1;\mu) \\
&\hspace{-2cm}+\frac{\vec{\gamma}_{\Delta^2_{ij}}}{2}  \left( \Delta_i^+(\mu) \Delta_j^-(\mu)+\Delta_i^-(\mu) \Delta_j^+(\mu) \right)   \,, \nonumber \\
\frac{\mathrm{d}}{\mathrm{d} \ln \mu^2} \vec{\sigma}^{\mathrm{asy}}(2;\mu) & =-\hat{\gamma}(3) \vec{\sigma}^{\mathrm{asy}}(2;\mu) \\
&+\vec{\gamma}_{\Delta^2_{ij}} \Delta^{\mathrm{asy}}_i(\mu)\Delta^{\mathrm{asy}}_j(\mu)\,,\nonumber
\end{align}
where $\sigma_i(2;\mu)=T_i(2;\mu)-T_i(1;\mu)^2$, $\Delta_i(\mu)=T_i(1;\mu)-T_g(1;\mu)$, and the joint central moments are defined as 
\begin{align}
&\sigma_i^{+-}(n,m;\mu) \\
&\hspace{-0.1cm}= \hspace{-0.1cm}\int_0^1\hspace{-0.2cm}  dx \,dy\, (x-T_i^+(1;\mu))^n\,(y-T_i^-(1;\mu))^m\, T_i^{+-}(x,y;\mu)\,. \nonumber
\end{align}

Here the vector is in flavor space, and $\gamma_{ij}^{(L)}(k)=-\int_0^1\df z\ z^{k-1}P_{ij}^{(L)}(z)$ is a moment of the timelike splitting functions \cite{Stratmann:1996hn,Mitov:2006ic,Moch:2007tx,Almasy:2011eq,Chen:2020uvt},  and $\vec{\gamma_{\Delta^2_{ij}}}$ is a new anomalous dimension that we have computed in perturbation theory.
Due to the positivity of the eigenvalues of the anomalous dimension matrix, $\hat{\gamma}(3)$,  \cite{Nachtmann:1973mr,Komargodski:2016gci}, the covariance of the joint track function, $\sigma^{+-}(1,1;\mu)$, decreases monotonically as it is evolved to the UV, as can be clearly seen in \Fig{fig:joint_track}.

\begin{figure}
\includegraphics[scale=0.28]{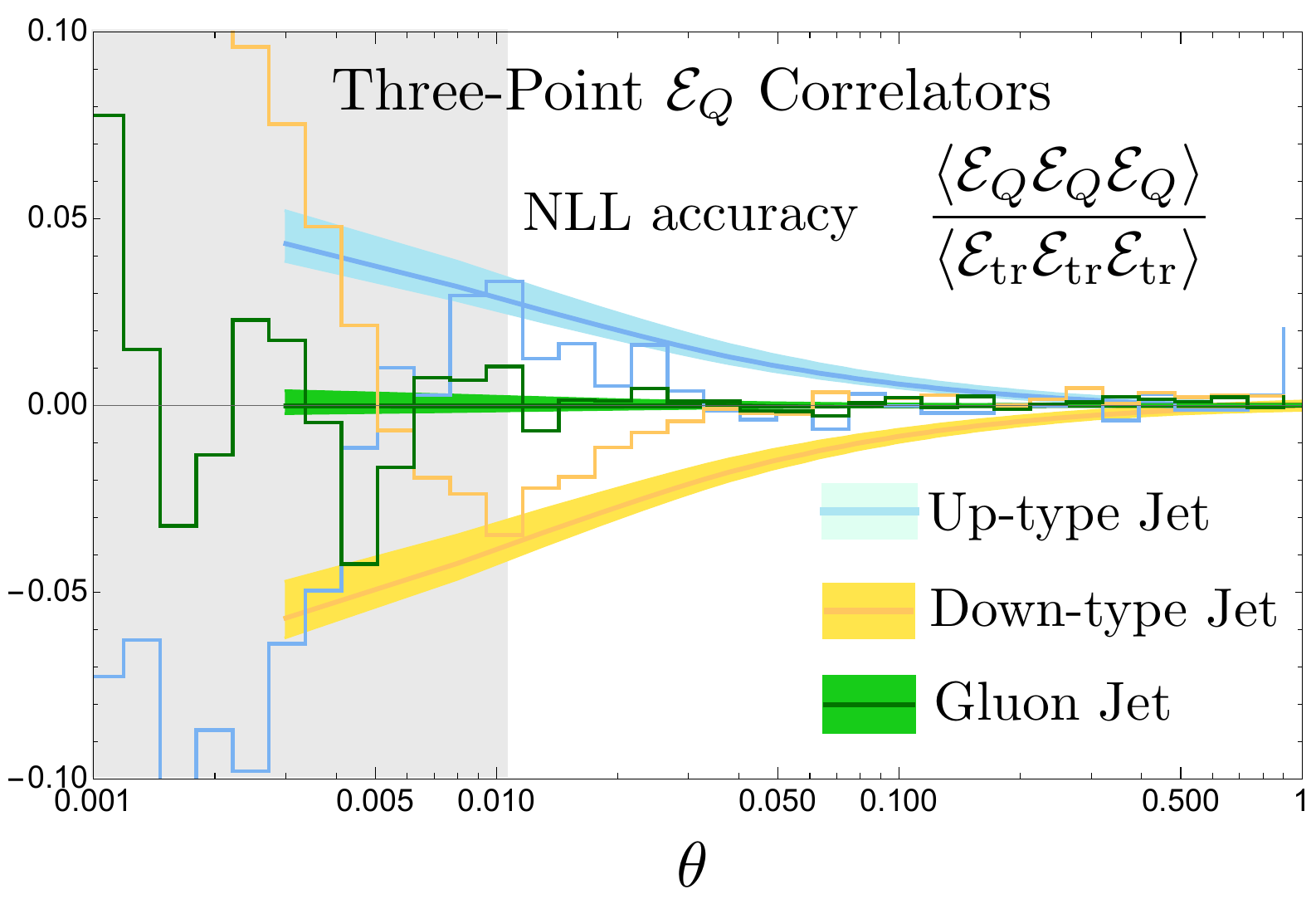} 
\caption{The scaling of the C-odd three-point correlator, $\langle \cE_\cQ \cE_\cQ \cE_\cQ\rangle$, for up-type and up-type quarks, and for gluons. Analytic results at NLL accuracy are shown in solid, and Pythia is shown as a histogram. The transition to the non-perturbative regime is shown in grey. $C$-invariance dictates that this distribution vanishes for gluons, and gives nearly opposite values for up-type and down-type quarks.}
\label{fig:EQ_threepoint}
\end{figure} 

\emph{Small Angle Scaling of Charged Correlators.}---We can apply the joint track function formalism to derive factorization formulas for multi-point correlators involving charge, allowing their scaling behavior in the small angle limit to be derived from the RG.  Following, \cite{Dixon:2019uzg} we work in terms of the cumulant, $\Sigma(z, \ln Q^2/\mu^2 , \mu)$ of the energy correlator,  expressed in terms of the variable $z=(1-\cos\theta)/2$, where $\theta$ is the angle between the detectors.  As for the standard energy correlator \cite{Dixon:2019uzg,Chen:2020vvp,Lee:2022ige}, or the energy correlator on tracks \cite{Jaarsma:2023ell}, we can perform a factorization into an inclusive hard function and a jet function incorporating the infrared measurement. In QCD, this factorization takes the form
\begin{align}
\label{eq:fact}
 \Sigma(z, \ln \frac{Q^2}{\mu^2} , \mu)
= \int_0^1 dx\, x^2 \vec{J} (\ln\frac{z x^2 Q^2}{\mu^2},\mu)
   \cdot  \vec{H} (x,\frac{Q^2}{\mu^2},\mu) \,,
\end{align}
where $\vec{J} = \{ J_u, J_{\bar u}, J_{d}, J_{\bar d},J_{s}, J_{\bar s},J_{c}, J_{\bar c}, J_g\}$ is a vector in flavor space, and similarly for $\vec{H}$. Note that as compared to the standard energy correlator, we must treat quarks and anti-quarks, as well as different flavors, distinctly. Here $Q$ is an appropriate hard scale, for example in the case of a jet at the LHC, $Q=p_T R$. Using the joint track functions, we can factorize the jet functions into a perturbative matching coefficient and (moments of) joint track functions.

This factorization allows us to use the renormalization group to derive the asymptotic scaling behavior of these more general correlators in the small angle limit. At leading logarithmic (LL) order, the jet functions take a particularly simple closed form, where we find for the two-point $\langle \cE_+(n_1) \cE_-(n_2) \rangle$, 
\begin{align}
\label{eq:chargeLL}
&\vec{J}_{\rm LL}^{+-}\left(\ln \frac{z Q^2}{\mu^2}, \mu\right)=
\vec{T}^{+-}(1,1;\sqrt{z}\,Q)
\cdot \left(\frac{\alpha_s(\sqrt{z}\,Q)}{\alpha_s(\mu)}\right)^{-\frac{\hat{\gamma}^{(0)}(3)}{\beta_0}}\,,
\end{align}
and for the two-point $\langle \cE_Q(n_1) \cE_Q(n_2) \rangle$, 
\begin{align}
\label{eq:chargeprodLL}
&\vec{J}_{\rm LL}^{~\cE_\cQ}\left(\ln \frac{z Q^2}{\mu^2}, \mu\right)=
\vec{T}^{\mathrm{asy}}(2;\sqrt{z}\,Q)
\cdot \left(\frac{\alpha_s(\sqrt{z}\,Q)}{\alpha_s(\mu)}\right)^{-\frac{\hat{\gamma}^{(0)}(3)}{\beta_0}}\,,
\end{align}
highlighting how different observables probe different moments of the joint track function. This generalizes results for the standard energy correlators \cite{Konishi:1979cb,Belitsky:2013ofa,Dixon:2019uzg,Kologlu:2019mfz,Korchemsky:2019nzm} to correlators with charge.

We have extracted the relevant moments of the joint track functions from Pythia \cite{Sjostrand:2014zea,Sjostrand:2007gs}, enabling predictions for the scaling of a number of generalized correlators. Although the analytic formulas above were presented at LL for simplicity, all numerical results will be presented at collinear NLL. Additionally, we find that it is essential to incorporate the leading non-perturbative power corrections. A key feature of the energy correlators is that the functional form of the leading power correction is fixed to scale as $\Lambda/\theta^3$, and is an additive contribution \cite{Belitsky:2001ij,Korchemsky:1999kt,Korchemsky:1997sy,Korchemsky:1995zm,Korchemsky:1994is,Schindler:2023cww}. This continues to hold when (joint) track functions are incorporated, with $\Lambda$ being universal up to multiplication by moments of the (joint) track functions \cite{Jaarsma:2023ell}. We observe a nontrivial interplay between different values of $\Lambda$ for different type of charge correlators.

In \Fig{fig:scaling_pm}, we show the small angle scaling of two-point correlations of the $\cE_+$ and $\cE_-$ detectors. Analytic results are shown in solid, and results from Pythia are shown in histograms. We see that confinement produces enhanced small angle correlations between like-sign hadrons relative to same-sign hadrons. It is crucial to emphasize that these correlations are due to QCD interactions,  \emph{not} electromagnetic interactions. 

In \Fig{fig:scaling_EQ} we show the two-point correlator $\langle \cE_\cQ \cE_\cQ \rangle$ for up-type quarks.   An interesting feature of correlations of the $\cE_{\cQ}$ detectors is that  they exhibit large cancellations between positive and negative contributions, leading to scaling laws that differ by an integer amount from $\cE$ correlators. As can be seen from \Fig{fig:scaling_EQ}, they scale like $d\sigma/d\theta \sim \theta^{-2}$ up to logarithmic corrections, instead of the familiar  $d\sigma/d\theta \sim \theta^{-1}$ of energy correlators. 

In \Fig{fig:EQ_threepoint} we consider the particularly interesting case of the scaling of the three-point correlator $\langle \cE_\cQ \cE_\cQ \cE_\cQ \rangle$, which is a $C$-odd observable. Results are shown for up-type quarks, down-type quarks, and gluons, with theoretical predictions in solid, and results from Pythia in histograms. This correlator is a $C$-odd observable,  giving precisely zero for gluons, and close to opposite results for up-type and down-type quarks due to approximate isospin symmetry. 

These examples illustrate the ability to extract scaling behavior for correlators involving charges in QCD for the first time. They highlight the theoretical advantages of reformulating jet substructure as the study of correlation functions: correlation functions are only sensitive to moments of (joint) track functions, and the scaling behavior of their leading non-perturbative power corrections is predicted. The charged correlators introduced here provide a broad generalization of the energy correlator program.

We have found that our results are sensitive to the non-perturbative inputs, highlighting the sensitivity of these observable to properties of the confinement transition, as expected. It would therefore be interesting to extract the joint track functions in different hadronization models, particularly comparing  the string  \cite{Andersson:1983ia} vs. cluster \cite{Marchesini:1991ch,Bahr:2008pv,Bahr:2008tf,Bellm:2019zci} models. This is beyond the scope of the current study.

\begin{figure}
\includegraphics[width=0.48\textwidth]{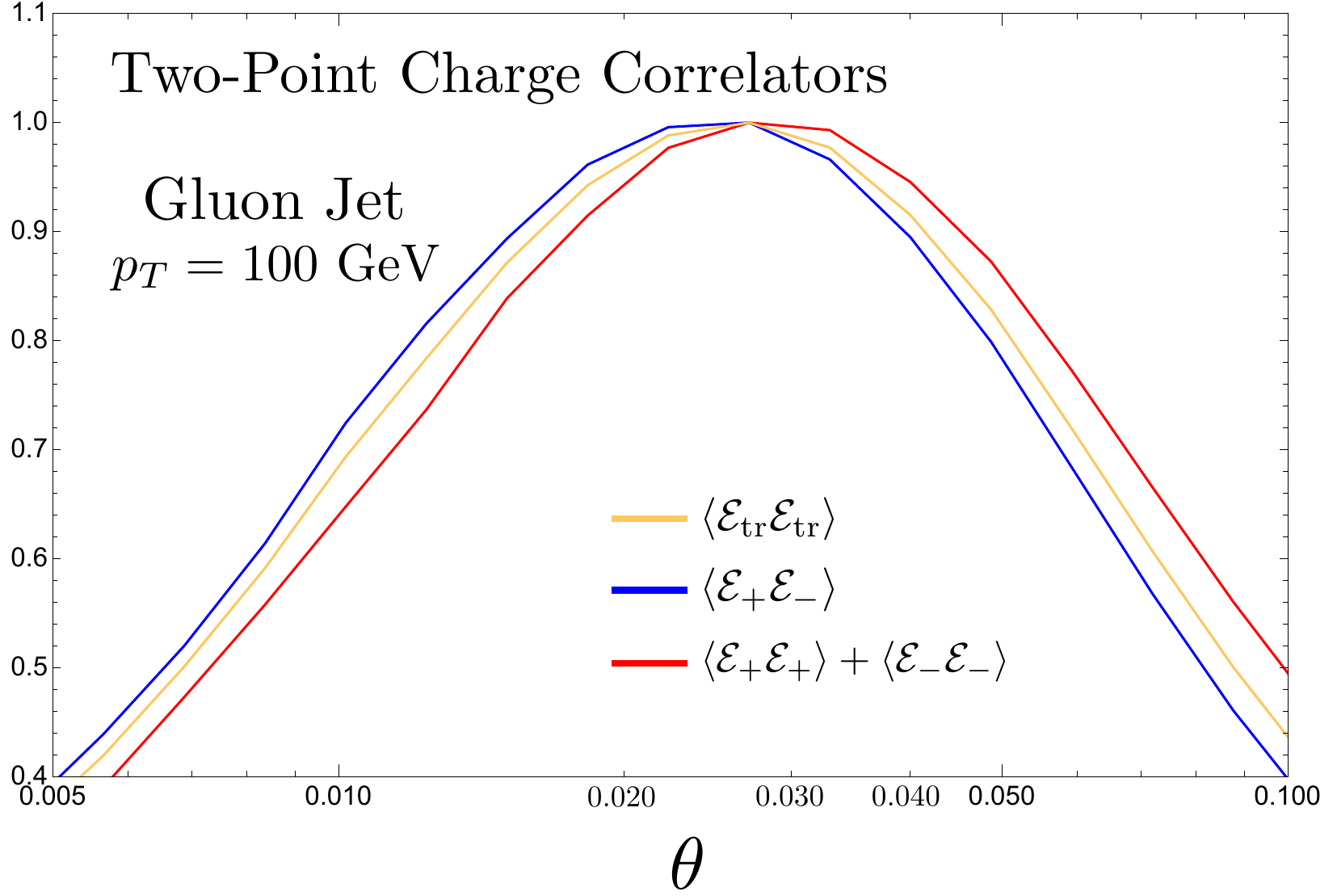}
  \caption{The confinement transition through the lens of different charged correlators, as computed using Pythia. The location and width of the confinement transition are modified depending on the quantum numbers correlated.}
  \label{fig:transition}
\end{figure}

\emph{A Refined View of the Hadronization Transition.}---As first illustrated in \cite{Komiske:2022enw}, a remarkable feature of the energy correlators is their ability to image the hadronization transition. This transition has since been measured by both ALICE and STAR \cite{talk1,talk2,talk3}. Using the $\langle \cE_+(n_1) \cE_-(n_2) \rangle$ and $\langle \cE_+(n_1) \cE_+(n_2) \rangle$, we can now study more refined features of the transition, tracking how  electric charge is confined.  

In \Fig{fig:transition} we show the confinement transition as probed using the standard energy correlators on tracks,  and the $\langle \cE_+(n_1) \cE_-(n_2) \rangle$ and $\langle \cE_+(n_1) \cE_+(n_2) \rangle$ charged correlators. We see that the transition is narrower for $\langle \cE_+(n_1) \cE_-(n_2) \rangle$ as compared to $\langle \cE_+(n_1) \cE_+(n_2) \rangle$, and that the peak occurs at different scales. We also observe abrupt changes in structure of the $C$-odd three-point correlator at the transition scale, see \Fig{fig:EQ_threepoint}. While we are not currently able to interpret these in terms of the underlying microscopic dynamics, we believe that charged correlators provide a new window into the confinement transition, motivating experimental measurements, and detailed comparisons with different hadronization models.

\emph{Conclusions.}---In this \emph{Letter} we have shown that the recent reformulation of jet substructure in terms of energy correlators can be extended to enable the calculation of a much broader class of observables, $\langle \cE_{R_1}(n_1) \cE_{R_2}(n_2) \cdots \cE_{R_k}(n_k) \rangle$, correlating the energy flux carried by hadrons of different quantum numbers. To compute these observables,  we introduced new non-perturbative functions characterizing the fragmentation process, the ``joint track functions", whose moments enter the factorization theorems for these observables.  

We applied the joint track functions to the two-point correlators, $\langle \cE_+(n_1) \cE_-(n_2) \rangle$, and $\langle \cE_+(n_1) \cE_+(n_2) \rangle$, to show that the strong interactions introduce enhanced small angle correlations between opposite-sign hadrons, relative to like-sign hadrons. We were also able to compute a $C$-odd three-point function $\langle \cE_\cQ \cE_\cQ \cE_\cQ \rangle$. 

Our calculations combine perturbation theory, factorization theorems and the renormalization group to greatly extend the class of calculable jet substructure observables. 
Building on recent measurements of the energy correlators \cite{talk1,talk2,talk3}, we look forward to the application of these new charged correlators to better understanding the dynamics of real world confinement using collider data. 

\emph{Acknowledgements.}---We thank Andrew Tamis, Helen Caines, Gregory Korchemsky, Berndt Mueller, Youqi Song for useful discussions and/or comments on the manuscript. K.L.~was supported by the U.S.~DOE under contract number DE-SC0011090.  I.M. is supported by start-up funds from Yale University.

\bibliography{EEC_ref.bib}{}
\bibliographystyle{apsrev4-1}
\newpage
\onecolumngrid
\newpage

\end{document}